\documentclass[conference]{IEEEtran}
\usepackage{amsmath}
\usepackage[pdftex]{graphicx}
\usepackage{graphicx}
\usepackage{pstricks}
\usepackage{amsfonts}
\usepackage{tabularx}

\usepackage{amssymb}
\usepackage{multirow}
\usepackage{epsfig}
\usepackage{epstopdf}

\usepackage[noadjust]{cite} 

\begin{document}
\title{Secrecy Outage Probability Analysis for Downlink NOMA with Imperfect SIC at Untrusted Users}
\author{\IEEEauthorblockN{Sapna Thapar$^{1}$, Insha Amin$^{1}$, Deepak Mishra$^{2}$, and Ravikant Saini$^{1}$}
\IEEEauthorblockA{$^{1}$Department of Electrical Engineering, Indian Institute of Technology Jammu, India\\
$^{2}$School of Electrical Engineering and Telecommunications, University of New South Wales, Australia\\
Emails: thaparsapna25@gmail.com, 2018ree0052@iitjammu.ac.in, d.mishra@unsw.edu.au, ravikant.saini@iitjammu.ac.in
}}
\maketitle

\begin{abstract}
Non-orthogonal multiple access (NOMA) has come
to the fore as a spectrally efficient technique for fifth-generation
networks and beyond. At the same time, NOMA faces severe
security issues in the presence of untrusted users due to successive
interference cancellation (SIC)-based decoding at receivers. In
this paper, to make the system model more realistic, we consider
the impact of imperfect SIC during the decoding process.
Assuming the downlink mode, we focus on designing a secure
NOMA communication protocol for the considered system model
with two untrusted users. In this regard, we obtain the power
allocation bounds to achieve a positive secrecy rate for both near
and far users. Analytical expressions of secrecy outage probability
(SOP) for both users are derived to analyze secrecy performance.
Closed-form approximations of SOPs are also provided to gain
analytical insights. Lastly, numerical results have been presented,
which validate the exactness of the analysis and reveal the effect
of various key parameters on achieved secrecy performance.
\end{abstract}

\section{Introduction}
Non-orthogonal multiple access (NOMA) has been perceived as a promising enabling technology for fifth-generation (5G) wireless networks and beyond, as it entertains the possibility of serving more users in limited available resources \cite{nomasurvey}. At the same time, the broadcast nature of NOMA poses a security problem as the signal is vulnerable to eavesdropping. The use of physical layer security (PLS) has sparked widespread interest in solving the security concerns of the information-carrying signal in wireless communication. Therefore, achieving secure NOMA communication by utilizing the potential of PLS is a promising area of research \cite{8509094}.

\subsection{Related Works}
Based on the concept of PLS, existing works have the prior objective of securing the information-carrying signal against external eavesdroppers \cite{8509094}. Additionally, the users who share the same resource block in NOMA may be untrusted, thereby making it compulsory to provide secrecy against internal eavesdropping. An untrusted users' scenario is a hostile situation where no users have mutual trust amongst each other, and therefore, they focus on securing their data from others \cite{saini2016ofdma}, \cite{9188014}. In this regard, \cite{basepaper}, \cite{7833022} have considered the secrecy issue of only near user against the far untrusted user. However, a robust NOMA system should be designed such that even the far user is provided secrecy against the near untrusted user. Based on this, \cite{9172088} proposed a PLS design for NOMA with a stronger near untrusted user. In \cite{9004475}, a directional demodulation approach is followed to protect the data of weak user from an untrusted strong user. In \cite{9169675}, a linear precoding technique is proposed to prevent NOMA users from eavesdropping on each other. In \cite{9188014}, \cite{globecom}, a novel secure decoding order is suggested to provide positive secrecy rate for both strong and weak users. In \cite{9324786}, feasible secure decoding orders are investigated to ensure a positive secrecy rate for all users in an $N$-user system.

\subsection{Research Gap and Motivation}
A common assumption in \cite{9172088}-\cite{globecom} is that perfect successive interference cancellation (SIC) is performed by the receivers. Here, the interference from the decoded users is cancelled altogether while decoding later users. However, this might not be a realistic approach due to practical implementation problems such as decoding errors and complexity scaling \cite{7881111}. Consequently, imperfect SIC, where the residual interference (RI) from incorrectly decoded users remains while decoding later users, would be a practical model \cite{7881111}, \cite{8755843}. In NOMA literature, researchers have assumed either a fixed value of RI \cite{constant_sic1}, \cite{constant_sic2} or considered RI as
a linear function of the interfering power \cite{7881111}, \cite{8755843}. However, in the direction of untrusted NOMA security, imperfect SIC has not received much focus yet. In \cite{9188014}, secrecy outage probability (SOP)  of an untrusted NOMA system has been analyzed with a fixed RI value, which is a strong unrealistic assumption. In contrast, the linear model can more effectively represent the relationship between RI and power of the received signal. In \cite{9324786}, though the linear SIC model has been considered, no SOP analysis has been done. \emph{Hence, to analyze the realistic impact of imperfect SIC, we investigate the secrecy performance analysis for a two-user untrusted NOMA system with linear SIC model, which to the best of our knowledge, has not been explored yet.}

\subsection{Key Contributions}
The key contributions of this work are summarized below: (1) Considering the impact of RI with linear imperfect SIC model in a two-user untrusted NOMA system, the power allocation (PA) bounds to achieve a positive secrecy rate for both users are investigated. (2) To analyze secrecy performance, the analytical expressions of SOP for both near and far users are derived. (3) The exact closed-form approximations of SOPs have also been obtained to attain analytical insights.
(4) Numerical results have been provided to validate the analytical expressions, followed by insightful discussions on the impact of different key parameters on the system performance.

\begin{figure}[!t]
\centering
\includegraphics[scale=.28]{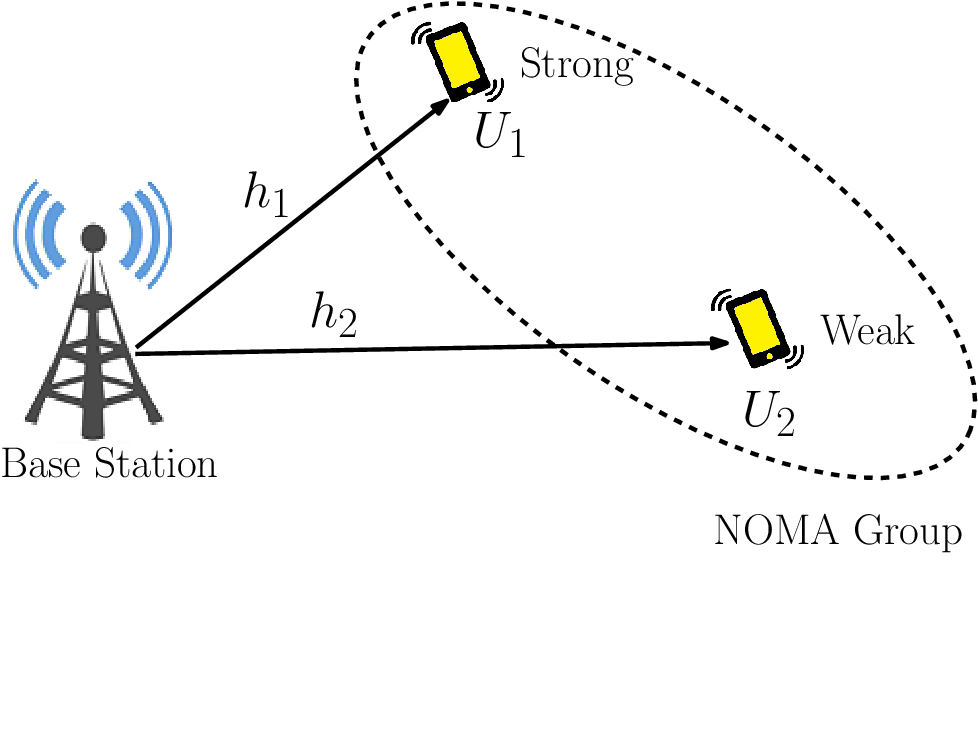}
\vspace{-1.1cm}
\caption{Illustration of a downlink two-user untrusted NOMA system.}
\label{SystemModel}
\end{figure}

\section{NOMA Transmission among Untrusted Users}
Here we first present the system model. Then, the possible decoding orders for an untrusted NOMA system are discussed. 
\subsection{System Model and NOMA Principle}
Downlink of a NOMA system is considered, where one base station (BS) communicates with two untrusted users (Fig. \ref{SystemModel}). We denote the $n$-th user by $U_{n}$, where $n\in \mathcal{N}=\{1, 2\}$. All the nodes in the network are assumed to be equipped with one antenna. The Rayleigh fading channel gain coefficient from BS to $U_{n}$ is denoted by  $h_{n}$. The channel power gains $|h_{n}|^{2}$ obeys an exponential distribution with mean parameter $\lambda_{n}=L_{p}d_{n}^{-e}$, where $d_{n}$ denotes the distance between BS and $U_{n}$, $L_{p}$ is the path loss constant, and $e$ indicates the path loss exponent.Without loss of generality, we assume that $|h_{1}|^{2}>|h_{2}|^{2}$, and thus, near and far user, i.e., $U_{1}$ and $U_{2}$ could be regarded as strong and weak user, respectively. The BS superposes information signals of users and broadcasts the superimposed signal with a total BS transmission power $P_{t}$. The fraction of $P_{t}$ allocated for $U_{1}$ is denoted by  $\alpha$. The remaining fraction $(1-\alpha)$ is alloted to $U_{2}$. At the receiver side, each user performs SIC wherein inter-user interference imposed by the superposition is cancelled out to extract the desired signal \cite{7343355}. Without loss of generality, received additive white Gaussian noise is assumed with mean $0$ and variance $\sigma^{2}$ at both users. We consider an imperfect SIC scenario where RI from inaccurately decoded signals exists while decoding later users. $\beta$, $0\leq\beta\leq1$, denotes the RI factor, where $\beta=1$ corresponds to the scenario of maximum interference, and $\beta=0$ indicates perfect SIC \cite{7881111}, \cite{8755843}.

\subsection{Decoding Orders for Untrusted NOMA}\label{decodingorders}
In an untrusted NOMA system, during the SIC process, each user can decode its own signal and other users' signal as well \cite{9188014}, \cite{7343355}. This SIC process is performed in a certain sequence, which is known as the ''\textit{decoding order}'' of the system. In a two users' scenario, the total possible decoding orders are $4$ \cite{9188014}. Let us denote the decoding order as a $2\times 2$ matrix $\mathbf{D}_{o}$, where $o \in \{1,2,3,4\}$ represents the index of $o$-th decoding order. Here $m$-th column of matrix $\mathbf{D}_{o}$ is specified by a column vector $\mathbf{d}_{m}$ of size $2\times1$, which depicts the SIC sequence observed by $U_{m}$, where $m \in \mathcal{N}$. To be more explicit, $[\mathbf{d}_{m}]_{k}=n$ signifies that $U_{m}$ decodes data of $U_{n}$ at $k$-th stage, where $n, k \in \mathcal{N}$ and $[\mathbf{d}_{m}]_{1} \neq [\mathbf{d}_{m}]_{2}$. 
Thus, the $4$ possible decoding orders can be written as $\mathbf{D}_{1}=[2,1;2,1]$, $\mathbf{D}_{2}=[2,1;1,2]$, $\mathbf{D}_{3}=[1,2;2,1]$, and $\mathbf{D}_{4}=[1,2;1,2]$. 
In \cite[Theorem 2]{9188014}, it is proved that the optimal decoding order with respect to providing maximum secrecy rate at both users is $\mathbf{D}_{2}$. So, all further investigations will be carried out for $\mathbf{D}_{2}$. %However, the same can be verified for other decoding orders as well. 

\section{PA Bounds from Secrecy Perspective}
With the objective of securing one user's data from another, first we investigate the feasible PA bounds that ensure positive secrecy rate for both users. In $\mathbf{D}_{2}$, both near and far users first decode signals of other user, perform SIC, and then decode their own signal \cite{9188014}. Thereby, using linear SIC model \cite{9324786}-\cite{8755843}, the achievable signal-to-interference-plus-noise-ratio (SINR) $\Gamma_{nm}$ at $U_{m}$, when $U_{n}$ is decoded by $U_{m}$,  where $m, n \in \mathcal{N}$, is given as 
\begin{align}
\Gamma_{21} = \frac{(1-\alpha)|h_{1}|^{2}}{\alpha|h_{1}|^{2}+\frac{1}{\rho_{t}}}, \quad &
\Gamma_{12} = \frac{\alpha|h_{2}|^{2}}{(1-\alpha)|h_{2}|^{2} + \frac{1}{\rho_{t}}},  \nonumber \\
\Gamma_{11} = \frac{\alpha|h_{1}|^{2}}{(1-\alpha)\beta|h_{1}|^{2}+\frac{1}{\rho_{t}}}, \quad &
\Gamma_{22} = \frac{(1-\alpha)|h_{2}|^{2}}{\alpha\beta|h_{2}|^{2}+\frac{1}{\rho_{t}}},
\end{align}
where $\rho_{t}\stackrel{\Delta}{=}\frac{P_{t}}{\sigma^{2}}$ denotes the BS transmit signal-to-noise ratio (SNR). The  achievable secrecy rates $R_{s1}$ and $R_{s2}$ at $U_{1}$ and $U_{2}$, respectively, can be given as \cite{8509094} 
\begin{equation}\label{secrecy_rate}
R_{s1} = R_{11} - R_{12}, \quad R_{s2} = R_{22} - R_{21},
\end{equation}
where $R_{nm} = \log_{2}(1+\Gamma_{nm})$ denotes the data rate at $U_{m}$ as given by Shannon's Theorem. To achieve positive secrecy rate for a given user, the rate of the legitimate channel has to be higher than that of the eavesdropper's channel. Thus, for $U_{1}$, the positive secrecy rate condition, i.e.,  $R_{11}> R_{12}$, which simplifies to $\Gamma_{11} > \Gamma_{12}$, must be appeased. This gives
\begin{equation}\label{alpha1}
\alpha < 1 + \frac{|h_{1}|^{2}-|h_{2}|^{2}}{|h_{1}|^{2}|h_{2}|^{2}\rho_{t}(1-\beta)}.
\end{equation}
Similarly, $\Gamma_{22}>\Gamma_{21}$ to obtain positive $R_{s2}$ for $U_{2}$ gives
\begin{equation}\label{alpha2}
\alpha >  \frac{|h_{1}|^{2}-|h_{2}|^{2}}{|h_{1}|^{2}|h_{2}|^{2}\rho_{t}(1-\beta)}.
\end{equation}
From \eqref{alpha1} and \eqref{alpha2}, we can easily infer that in decoding order $\mathbf{D}_{2}$, positive secrecy rate can be obtained at both the users, provided $\frac{|h_{1}|^{2}-|h_{2}|^{2}}{|h_{1}|^{2}|h_{2}|^{2}\rho_{t}(1-\beta)} < \alpha < 1 + \frac{|h_{1}|^{2}-|h_{2}|^{2}}{|h_{1}|^{2}|h_{2}|^{2}\rho_{t}(1-\beta)}$.

\begin{figure*}[t!]
\begin{align}\label{so1}
s_{1} &= \text{Pr}\{R_{s1} <  R_{s1}^{th}\}\!=\!\text{Pr}\bigg\{\frac{1+\Gamma_{11}}{1+\Gamma_{12}} < \Pi_{1}\bigg\}\!=\!\text{Pr}\bigg\{|h_{1}|^{2} D_{1} < N_{1} \bigg\} =\text{Pr}\bigg\{|h_{1}|^{2} < \frac{N_{1}}{D_{1}}, D_{1}>0\bigg\}\!+\!\text{Pr}\bigg\{|h_{1}|^{2} \geq \frac{N_{1}}{D_{1}}, D_{1}\leq 0\bigg\}, \nonumber \\
&=\text{Pr}\bigg\{|h_{1}|^{2} < \frac{N_{1}}{D_{1}}, |h_{2}|^{2} < T_{1} \bigg\} + \text{Pr}\bigg\{|h_{1}|^{2} \geq \frac{N_{1}}{D_{1}}, |h_{2}|^{2} \geq T_{1} \bigg\}, \nonumber \\
&=
\begin{cases}
 \int_{0}^{T_{1}} \bigg(1-\exp\bigg\{\frac{-N_{1}}{D_{1}\lambda_{1}}\bigg\}\bigg)f_{| h_{2} |^{2}}(y_{1}) dy_{1} + \int_{T_{1}}^{\infty} 1 \times f_{| h_{2} |^{2}}(y_{1}) dy_{1} , &  \alpha_{1a} < \alpha < \alpha_{1b} \nonumber \\
  \int_{0}^{\infty} \bigg(1-\exp\bigg\{\frac{-N_{1}}{D_{1}\lambda_{1}}\bigg\}\bigg)f_{| h_{2} |^{2}}(y_{1}) dy_{1},  &  \alpha \geq \alpha_{1b} \nonumber \\
\int_{0}^{\infty} 1 \times f_{| h_{2} |^{2}}(y_{1}) dy_{1} , & \text{otherwise} \nonumber 
\end{cases}
\\
&=
\begin{cases}
1 - \frac{1}{\lambda_{2}} \int_{0}^{T_{1}} \exp\bigg\{\frac{-((\Pi_{1}-1)((1-\alpha)y_{1}\rho_{t}+1)+\alpha y_{1} \rho_{t}\Pi_{1})}{(\alpha\rho_{t}((1-\alpha)y_{1}\rho_{t}+1)-(\Pi_{1}-1)((1-\alpha)y_{1}\rho_{t} + 1)\beta(1-\alpha)\rho_{t}-\Pi_{1}\beta\alpha(1-\alpha)\rho_{t}^{2}y_{1})\lambda_{1}} - \frac{y_{1}}{\lambda_{2}}\bigg\} dy_{1},   & \alpha_{1a} < \alpha < \alpha_{1b}  \\
1 - \frac{1}{\lambda_{2}} \int_{0}^{\infty} \exp\bigg\{\frac{-((\Pi_{1}-1)((1-\alpha)y_{1}\rho_{t}+1)+\alpha y_{1} \rho_{t}\Pi_{1})}{(\alpha\rho_{t}((1-\alpha)y_{1}\rho_{t}+1)-(\Pi_{1}-1)((1-\alpha)y_{1}\rho_{t} + 1)\beta(1-\alpha)\rho_{t}-\Pi_{1}\beta\alpha(1-\alpha)\rho_{t}^{2}y_{1})\lambda_{1}} - \frac{y_{1}}{\lambda_{2}}\bigg\} dy_{1}, &  \alpha \geq \alpha_{1b}  \\
1,  & \text{otherwise}
\end{cases}
\end{align}
\vspace{-0.3cm}
\noindent\rule{18cm}{0.5pt}
\end{figure*}

\begin{figure*}[t!]
\begin{align}\label{so2}
s_{2} &= \text{Pr}\{R_{s2} <  R_{s2}^{th}\}=\text{Pr}\Big\{\frac{1+\Gamma_{22}}{1+\Gamma_{21}} < \Pi_{2}\Big\}=\text{Pr}\bigg\{|h_{2}|^{2} D_{2} < N_{2} \bigg\}\!=\!\text{Pr}\bigg\{|h_{2}|^{2} < \frac{N_{2}}{D_{2}}, D_{2}>0\bigg\} + \text{Pr}\bigg\{|h_{2}|^{2} \geq \frac{N_{2}}{D_{2}}, D_{2}\leq 0\bigg\}, \nonumber \\
&=\text{Pr}\bigg\{|h_{2}|^{2} < \frac{N_{2}}{D_{2}}, |h_{1}|^{2} < T_{2} \bigg\} + \text{Pr}\bigg\{|h_{2}|^{2} \geq \frac{N_{2}}{D_{2}}, |h_{1}|^{2} \geq T_{2} \bigg\}, \nonumber \\
&=
\begin{cases}
 \int_{0}^{T_{2}} \bigg(1-\exp\bigg\{\frac{-N_{2}}{D_{2}\lambda_{2}}\bigg\}\bigg)f_{| h_{1} |^{2}}(y_{2}) dy_{2} + \int_{T_{2}}^{\infty} 1 \times f_{| h_{1} |^{2}}(y_{2}) dy_{2} , \quad &  \alpha_{2a} < \alpha < \alpha_{2b} \nonumber \\
  \int_{0}^{\infty} \bigg(1-\exp\bigg\{\frac{-N_{2}}{D_{2}\lambda_{2}}\bigg\}\bigg)f_{| h_{1} |^{2}}(y_{2}) dy_{2}, \quad \quad \quad \quad  \quad \quad  \quad  \quad &  \alpha \leq \alpha_{2a} \nonumber \\
\int_{0}^{\infty} 1 \times f_{| h_{1} |^{2}}(y_{2}) dy_{2} , & \text{otherwise} \nonumber 
\end{cases}
\\
&=
\begin{cases}
1- \frac{1}{\lambda_{1}} \int_{0}^{T_{2}} \exp\bigg\{\frac{-((\Pi_{2}-1)(\alpha y_{2}\rho_{t}+1)+(1-\alpha)y_{2}\rho_{t}\Pi_{2})}{((1-\alpha)\rho_{t}(\alpha y_{2}\rho_{t}+1)-(\Pi_{2}-1)(\alpha y_{2}\rho_{t} + 1)\beta\alpha\rho_{t}-\Pi_{2}\beta\alpha(1-\alpha)\rho_{t}^{2}y_{2})\lambda_{2}} - \frac{y_{2}}{\lambda_{1}}\bigg\} dy_{2}, \quad  & \alpha_{2a} < \alpha < \alpha_{2b}  \\
1 - \frac{1}{\lambda_{1}} \int_{0}^{\infty} \exp\bigg\{\frac{-((\Pi_{2}-1)(\alpha y_{2}\rho_{t}+1)+(1-\alpha)y_{2}\rho_{t}\Pi_{2})}{((1-\alpha)\rho_{t}(\alpha y_{2} \rho_{t}+1)-(\Pi_{2}-1)(\alpha y_{2}\rho_{t} + 1)\beta\alpha\rho_{t}-\Pi_{2}\beta\alpha(1-\alpha)\rho_{t}^{2} y_{2})\lambda_{2}} -  \frac{y_{2}}{\lambda_{1}}\bigg\} dy_{2}, & \alpha \leq \alpha_{2a} \\
1,  & \text{otherwise}
\end{cases}
\end{align}
\vspace{-0.3cm}
\noindent\rule{18cm}{0.5pt}
\end{figure*}

\begin{figure*}[t!]
\begin{align}\label{so1_C}
s_{1[C]}\!=\!\! 
\begin{cases}
1\!-\!\frac{1}{\lambda_{2}} \int_{0}^{T_{1[C]}} \exp\bigg\{\frac{-((\Pi_{1}-1)(\beta(1-\alpha)y_{1}\rho_{t} + 1)+ \alpha y_{1} \rho_{t}\Pi_{1})}{(\alpha\rho_{t}(\beta(1-\alpha)y_{1}\rho_{t}+1)-(\Pi_{1}-1)(\beta(1-\alpha)y_{1}\rho_{t} + 1)\beta(1-\alpha)\rho_{t}-\Pi_{1}\beta\alpha(1-\alpha)\rho_{t}^{2}y_{1})\lambda_{1}} - \frac{y_{1}}{\lambda_{2}}\bigg\} dy_{1}, \quad  & \alpha > \alpha_{1[C]} \\
1,  & \text{otherwise}
\end{cases}
\end{align}
\vspace{-0.3cm}
\noindent\rule{18cm}{0.5pt}
\end{figure*}

\begin{figure*}[t!]
\begin{align}\label{so2_C}
s_{2[C]} = \begin{cases}
1-\frac{1}{\lambda_{1}} \int_{0}^{T_{2[C]}} \exp\bigg\{\frac{-((\Pi_{2}-1)(\alpha y_{2}\rho_{t} + 1)+ (1-\alpha) y_{2} \rho_{t}\Pi_{2})}{((1-\alpha)\rho_{t}( \alpha y_{2}\rho_{t}+1)-(\Pi_{2}-1)(\alpha y_{2}\rho_{t} + 1) \alpha\rho_{t}-\Pi_{2}\alpha(1-\alpha)\rho_{t}^{2}y_{2})\lambda_{2}} - \frac{y_{2}}{\lambda_{1}}\bigg\} dy_{2}, \quad  & \alpha < \alpha_{2[C]} \\
1,  & \text{otherwise}
\end{cases}
\end{align}
\vspace{-0.3cm}
\noindent\rule{18cm}{0.5pt}
\end{figure*}

\begin{figure*}
\begin{align}\label{so1_asy}
\widehat s_{1} &= 
\begin{cases}
1-\bigg(\big(1-\exp\big\{\frac{-T_{1}}{\lambda_{2}}\big\}\big)\times\exp\bigg\{ \frac{-((\Pi_{1}-1)(1-\alpha)\rho_{t} + \alpha\rho_{t}\Pi_{1})}{(\alpha\rho_{t}^{2}(1-\alpha) - (\Pi_{1}-1)(1-\alpha)^{2}\beta\rho_{t}^{2} - \Pi_{1}\beta\alpha(1-\alpha)\rho_{t}^{2})\lambda_{1}}\bigg\}\bigg), \quad & \alpha_{1a} < \alpha < \alpha_{1b} \\ 
1-\exp\bigg\{ \frac{-((\Pi_{1}-1)(1-\alpha)\rho_{t} + \alpha\rho_{t}\Pi_{1})}{(\alpha\rho_{t}^{2}(1-\alpha) - (\Pi_{1}-1)(1-\alpha)^{2}\beta\rho_{t}^{2} - \Pi_{1}\beta\alpha(1-\alpha)\rho_{t}^{2})\lambda_{1}}\bigg\}, & \alpha \geq \alpha_{1b} \\
1, & \text{otherwise}
\end{cases}
\end{align}
\vspace{-0.3cm}
\noindent\rule{18cm}{0.5pt}
\end{figure*}

\begin{figure*}
\begin{align}\label{so2_asy}
\widehat s_{2} &= \begin{cases}
1-\bigg(\big(1-\exp\big\{\frac{-T_{2}}{\lambda_{1}}\big\}\big)\times\exp\bigg\{\frac{-((\Pi_{2}-1)\alpha\rho_{t} + (1-\alpha)\rho_{t}\Pi_{2})}{((1-\alpha)\rho_{t}^{2}\alpha - (\Pi_{2}-1)\alpha^{2}\rho_{t}^{2}\beta - (1-\alpha)\Pi_{2}\rho_{t}^{2}\alpha\beta)\lambda_{2}}\bigg\}\bigg), \quad & \alpha_{2a} < \alpha < \alpha_{2b} \\
1-\exp\bigg\{\frac{-((\Pi_{2}-1)\alpha\rho_{t} + (1-\alpha)\rho_{t}\Pi_{2})}{((1-\alpha)\rho_{t}^{2}\alpha - (\Pi_{2}-1)\alpha^{2}\rho_{t}^{2}\beta - (1-\alpha)\Pi_{2}\rho_{t}^{2}\alpha\beta)\lambda_{2}}\bigg\}, & \alpha \leq \alpha_{2a} \\
1, & \text{otherwise}
\end{cases}
\end{align}
\vspace{-0.3cm}
\noindent\rule{18cm}{0.5pt}
\end{figure*}

\begin{figure*}
\begin{align}\label{so1_C_asy}
\widehat s_{1[C]} &= \begin{cases}
1-\bigg(\big(1-\exp\big\{\frac{-T_{1[C]}}{\lambda_{2}}\big\}\big)\times\exp\bigg\{ \frac{-((\Pi_{1}-1)\beta(1-\alpha)\rho_{t} + \alpha\rho_{t}\Pi_{1})}{(\beta\alpha\rho_{t}^{2}(1-\alpha) - (\Pi_{1}-1)(1-\alpha)^{2}\beta^{2} \rho_{t}^{2}- \Pi_{1}\beta \alpha(1-\alpha)\rho_{t}^{2})\lambda_{1}}\bigg\}\bigg), \quad & \alpha > \alpha_{1[C]} \\
1, & \text{otherwise}
\end{cases}
\end{align}
\vspace{-0.3cm}
\noindent\rule{18cm}{0.5pt}
\end{figure*}

\begin{figure*}
\begin{align}\label{so2_C_asy}
\widehat s_{2[C]} &= \begin{cases}
1-\bigg(\big(1-\exp\big\{\frac{-T_{2[C]}}{\lambda_{1}}\big\}\big)\times\exp\bigg\{ \frac{-((\Pi_{2}-1) \alpha\rho_{t} + (1-\alpha)\rho_{t}\Pi_{2})}{(\alpha\rho_{t}^{2}(1-\alpha) - (\Pi_{2}-1)\alpha^{2} \rho_{t}^{2}- \Pi_{2} \alpha(1-\alpha)\rho_{t}^{2})\lambda_{2}}\bigg\}\bigg), \quad & \alpha < \alpha_{2[C]} \\
1, & \text{otherwise}
\end{cases}
\end{align}
\vspace{-0.4cm}
\noindent\rule{18cm}{0.5pt}
\end{figure*}

\section{Secrecy Performance Analaysis}\label{sectionIV}
In this section, we derive the analytical expressions of SOP to explore secrecy performance for $\mathbf{D}_{2}$. Asymptotic approximations are also provided to gain analytical insights.
\subsection{Secrecy Outage Probability}
The SOP is defined as the probability that the maximum achievable secrecy rate at a user is less than a threshold secrecy rate \cite{basepaper}. Let us denote SOP for $U_{n}$ as $s_{n}$, where $n \in \mathcal{N}$. 
\subsubsection{Near User} Assuming target and achievable secrecy rate, respectively, for $U_{1}$ as $R_{s1}^{th}$ and $R_{s1}$, the SOP $s_{1}$ is given as \eqref{so1}. Here $\text{Pr}\{.\}$ denotes the probability measure, $\Pi_{1} \stackrel{\Delta}{=} 2^{R_{s1}^{th}}$, $N_{1}=(\Pi_{1}-1)((1-\alpha)|h_{2}|^{2}\rho_{t}+1)+\alpha|h_{2}|^{2}\rho_{t}\Pi_{1}$, $D_{1}=\alpha\rho_{t}((1-\alpha)|h_{2}|^{2}\rho_{t}+1)-(\Pi_{1}-1)((1-\alpha)|h_{2}|^{2}\rho_{t} + 1)\beta(1-\alpha)\rho_{t}-\Pi_{1}\beta\alpha(1-\alpha)\rho_{t}^{2}|h_{2}|^{2}$, $T_{1}=\frac{\alpha-(\Pi_{1}-1)\beta(1-\alpha)}{(1-\alpha)\rho_{t}((\Pi_{1}-1)\beta(1-\alpha)+\alpha\Pi_{1}\beta-\alpha)}$, $\alpha_{1a}=\frac{\beta(\Pi_{1}-1)}{1+\beta(\Pi_{1}-1)}$, $\alpha_{1b}=\frac{\beta(\Pi_{1}-1)}{1-\beta}$, and $f_{\mid h_{2} \mid ^{2}}(x)$ is the probability density function (PDF) of $|h_{2}|^{2}$. Note that $\alpha_{1b}>\alpha_{1a}$, as this simplifies to $\beta\Pi_{1}>0$, which is always true. 

\subsubsection{Far user} 
Considering $R_{s2}^{th}$ and $R_{s2}$, respectively, as target and achievable secrecy rate, of $U_{2}$, $s_{2}$ is given as \eqref{so2}, where $\Pi_{2} \stackrel{\Delta}{=} 2^{R_{s2}^{th}}$, $N_{2}=(\Pi_{2}-1)(\alpha|h_{1}|^{2}\rho_{t}+1)+(1-\alpha)|h_{1}|^{2}\rho_{t}\Pi_{2}$, $D_{2}=(1-\alpha)\rho_{t}(\alpha|h_{1}|^{2}\rho_{t}+1)-(\Pi_{2}-1)(\alpha|h_{1}|^{2}\rho_{t} + 1)\beta\alpha\rho_{t}-\Pi_{2}\beta\alpha(1-\alpha)\rho_{t}^{2}|h_{1}|^{2}$, $T_{2}=\frac{(1-\alpha)-(\Pi_{2}-1)\beta\alpha}{\alpha\rho_{t}((\Pi_{2}-1)\beta\alpha+(1-\alpha)\Pi_{2}\beta-(1-\alpha))}$, $\alpha_{2a}=\frac{1-\Pi_{2}\beta}{1-\beta}$, $\alpha_{2b}=\frac{1}{1+\beta(\Pi_{2}-1)}$, and $f_{|h_{1}|^{2}}(x)$ is PDF of $|h_{1}|^{2}$. Here $\alpha_{2b}>\alpha_{2a}$, as it gives $\Pi_{2}>1$, which holds true. 

Similarly, we can obtain SOP expressions for other decoding orders. As a special case, the SOPs, $s_{1[C]}$ and $s_{2[C]}$ for conventional decoding order $\mathbf{D}_{1}$, are given in \eqref{so1_C} and \eqref{so2_C}, respectively. Here  $T_{1[C]}=\frac{\alpha-(\Pi_{1}-1)\beta(1-\alpha)}{(1-\alpha)\rho_{t}\beta((\Pi_{1}-1)\beta(1-\alpha)+\alpha(\Pi_{1}-1))}$, $\alpha_{1[C]}=\frac{(\Pi_{1}-1)\beta}{1+\beta(\Pi_{1}-1)}$, $T_{2[C]}=\frac{(1-\alpha)-(\Pi_{2}-1)\alpha}{\alpha\rho_{t}((\Pi_{2}-1)\alpha+(1-\alpha)(\Pi_{2}-1))}$ and $\alpha_{2[C]}=\frac{1}{\Pi_{2}}$. $[C]$ stands for conventional decoding order.

\subsection{Asymptotic Approximations}
Next, to provide analytical insights, we present closed-form approximations of SOPs at both users $U_{1}$ and $U_{2}$ for $\mathbf{D}_{2}$.

\subsubsection{Near User}
The exact closed-form expression of $s_{1}$, i.e., $\widehat s_{1}$, obtained by using $((1-\alpha)\rho_{t} y_{1}+1)$ $\approx$ $(1-\alpha)\rho_{t} y_{1}$ for $\rho_{t}\gg1$ in \eqref{so1} is given in \eqref{so1_asy}.
\subsubsection{Far User}
The closed-form asymptotic approximation $\widehat s_{2}$ of $s_{2}$, which we obtain by setting $(\alpha\rho_{t} y_{2}+1)$ $\approx$ $\alpha\rho_{t} y_{2}$ for high $\rho_{t}$ in \eqref{so2}, is provided in \eqref{so2_asy}.

Similar to above approximations, closed-form SOP expressions for other decoding orders can also be obtained. For conventional decoding order $\mathbf{D}_{1}$, the asymptotic SOP expressions, $\widehat s_{1[C]}$ and $\widehat s_{2[C]}$, are given in \eqref{so1_C_asy} and \eqref{so2_C_asy}, respectively. 

\section{Numerical Results}
Downlink NOMA system is considered with one BS and two untrusted users. Near user is assumed to be at a distance of $d_{1}=50$ meter from BS, and for far user, distance $d_{2}=100$ meter is adopted. Noise power is set to $-90$ dBm with noise signal following Gaussian distribution at all users. Small scale fading is presumed to obey an exponential distribution with a $1$ mean value at both links \cite{basepaper}. Simulations are averaged over $10^6$ randomly generated channel realizations by using Rayleigh distribution for both users. The value of $L_{p}$ and $e$, respectively, are taken to be $1$ and $3$. $\rho_{r}$ is assumed as the received SNR in decibels (dB) at $U_{2}$. The value of $\beta$ is taken to be $0.1$. Simulation, analytical, and asymptotic results are, respectively, marked as ‘Sim’, ‘Ana’, and ‘Asy’.

\begin{figure}[!t]
\centering
\includegraphics[scale=.38]{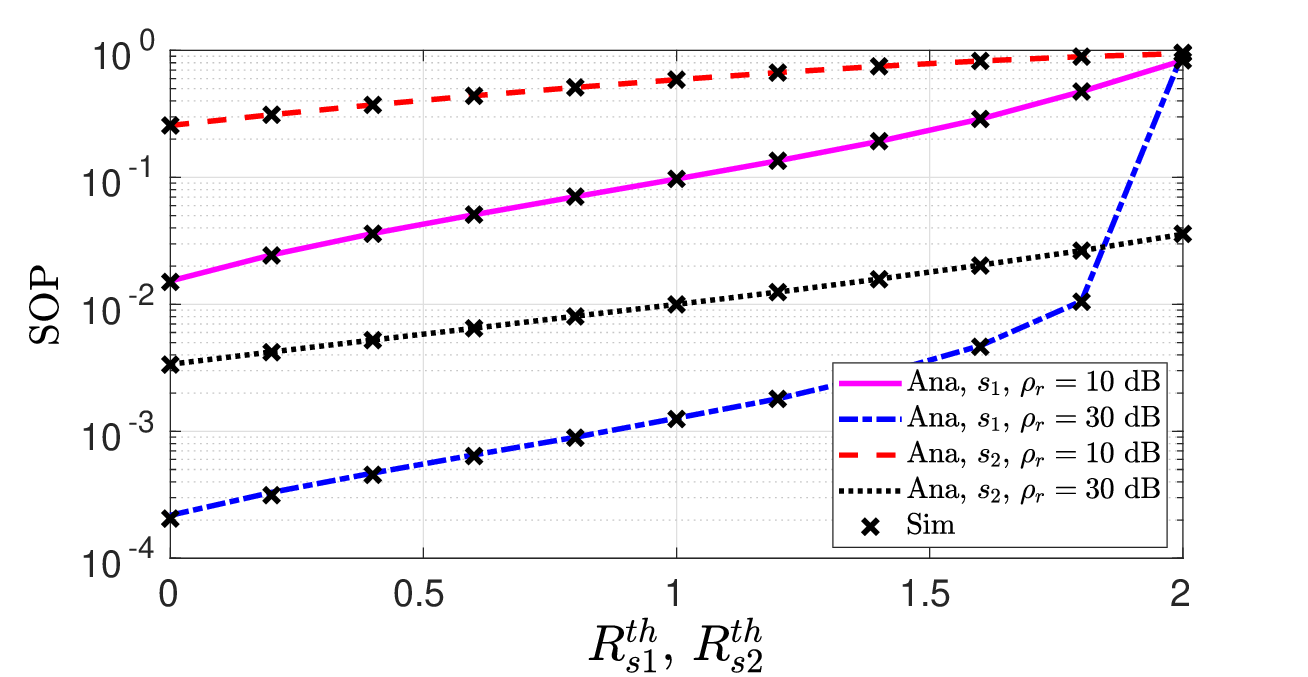}
\vspace{-0.1cm}
\caption{Validation of SOPs at $U_{1}$ and $U_{2}$ for $\mathbf{D}_{2}$, $\alpha=0.33$.}
\label{validation_sim_ana}
\end{figure}

\begin{figure}[!t]
\centering
\includegraphics[scale=.38]{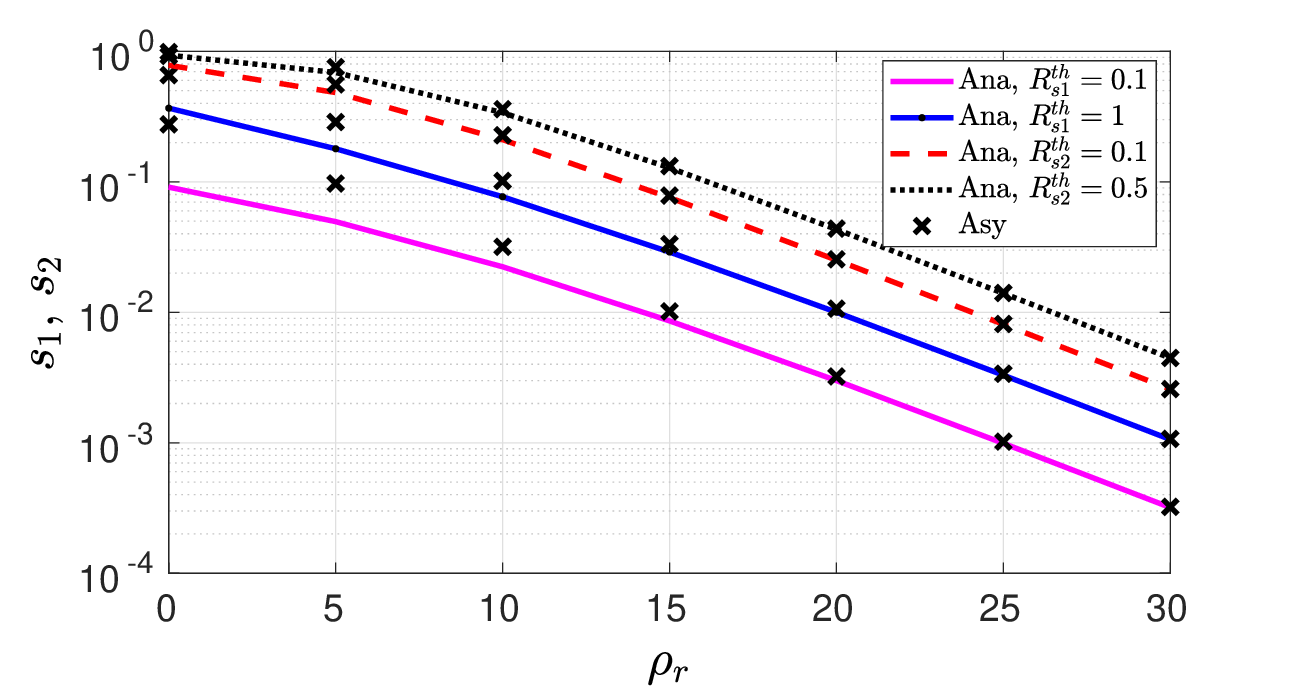}
\vspace{-0.1cm}
\caption{Validation of the accuracy of the proposed closed-form asymptotic approximations of SOPs for $\mathbf{D}_{2}$ with $\alpha=0.5$.}
\label{validation_ana_asy}
\end{figure}

\subsection{Validation of Analysis}
Here in Fig. \ref{validation_sim_ana}, the validation of SOPs, $s_{1}$ with $R_{s1}^{th}$ and $s_{2}$ with $R_{s2}^{th}$ for different values of $\rho_{r}$ are shown. The perfect agreement between simulated and analytical curves confirms the exactness of $s_{1}$ and $s_{2}$ analysis. It can be visualized from the results that $s_{1}$ and $s_{2}$ increase with the increase in threshold rates $R_{s1}^{th}$ and $R_{s2}^{th}$, respectively. Because outage occurs when the maximum achievable secrecy rate drops below a threshold rate, it is clear that increasing threshold secrecy rates at the user will, in turn, increase SOP. It can also be observed that an increase in $\rho_{r}$ decreases both $s_{1}$ and $s_{2}$. This happens because the secrecy rates achieved at users increase by an increase in SNR, and so, for a given threshold secrecy rate, SOP decreases. From Fig. \ref{validation_ana_asy}, we can  visualize that analytical results match with asymptotic results at high SNR, and it confirms the exactness of asymptotic expressions. 

\begin{figure}[!t]
\centering
\includegraphics[scale=.38]{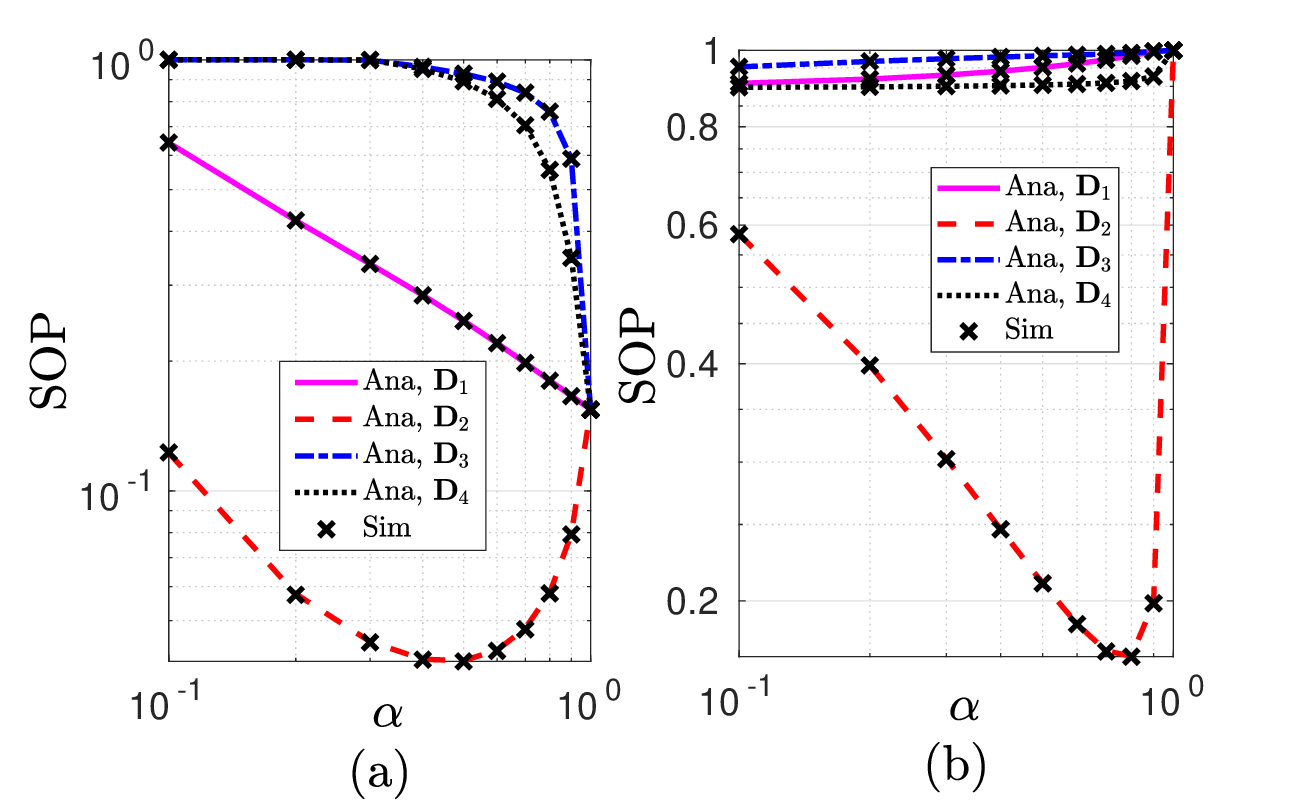}
%\vspace{-0.2cm}
\caption{Verification of the optimality of the decoding order $\mathbf{D}_{2}$ among $4$ possibilities \{$\mathbf{D}_{1}$, $\mathbf{D}_{2}$, $\mathbf{D}_{3}$, $\mathbf{D}_{4}$\}, for $U_{1}$ in (a) and $U_{2}$ in (b) with $\rho_{r}=10$ dB, $R_{s1}^{th}=0.5$ and $R_{s2}^{th}=0.1$.}
\label{validation_optimal_order_B}
\end{figure}

\begin{figure}[!t]
\centering
\includegraphics[scale=.39]{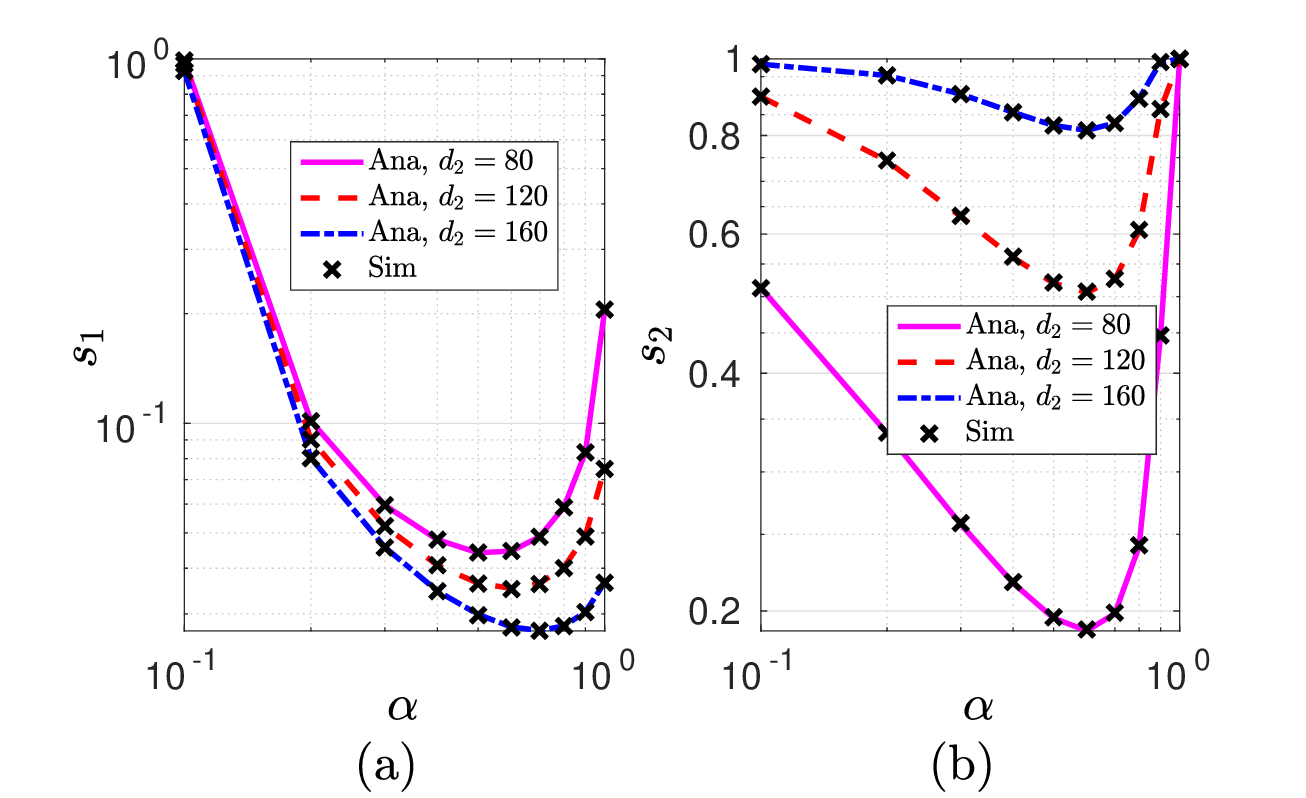}
\caption{Insights on optimal power allocation $\alpha$ that minimizes SOPs, $s_1$ in (a) and $s_2$ in (b), for different values of $d_{2}$, for $\mathbf{D}_{2}$, $d_{1}=40$ meter, $\rho_{t}=70$ dB, $R_{s1}^{th}=1$ and $R_{s2}^{th}=0.5$.}
\label{validation_optimal_PA}
\end{figure}

\subsection{Validation of Optimal Decoding Order}
Here we have sketched a plot to validate that the optimal decoding order for providing the highest secrecy rate for both near and far users is $\mathbf{D}_{2}$, as given in \cite[Theorem 2]{9188014}. In Fig. \ref{validation_optimal_order_B}, SOPs at both $U_{1}$ and $U_{2}$ for all $4$ decoding orders with respect to $\alpha$ are shown. The results corroborate that better SOP performance is obtained for $\mathbf{D}_{2}$, and hence it is optimal. 

\subsection{Impact of Relative Distance between Users}
In Fig. \ref{validation_optimal_PA}(a), we notice the effect of varying  the distance $d_{2}$ from BS on achievable SOPs. $d_{1}$ is fixed at $40$ meters. It can be seen that $s_{1}$ decreases with an increase in $d_{2}$. This happens because an increase in distance $d_{2}$ causes a drop in achievable data rate at $U_{2}$, which in turn provide better secrecy rate at $U_{1}$, thereby decreasing the SOP at $U_{1}$. Conversely, as shown in Fig. \ref{validation_optimal_PA}(b), a decrease in data rate at $U_{2}$ signifies a reduction in secrecy rate at $U_{2}$, which further increases the SOP for $U_{2}$. Thus, it can be observed that increasing the distance from BS to $U_{2}$ has a contradictory effect on $s_{1}$ and $s_{2}$. Fig. \ref{validation_optimal_PA}(a) and Fig. \ref{validation_optimal_PA}(b) also confirm the existence of an optimal PA that minimizes the SOP performance for both users $U_{1}$ and $U_{2}$.

\section{Concluding Remarks}
We have focused on the practical but adverse problem of SIC being imperfect in a secure NOMA system. Considering RI, the PA bounds are calculated to provide a positive secrecy rate for users. Analytical and asymptotic expressions of SOP are derived. Numerical results are provided to validate the analytical expressions and exhibit the effects of different key parameters on performance. Future work includes extending the study of SOP in a multi-user untrusted NOMA scenario.

\section*{Acknowledgement}
This work was partially supported by the Faculty Funded Postdoctoral Research Fellowship support of EET School at UNSW Sydney.

\bibliographystyle{IEEEtran}
\bibliography{ref}
\end{document}